# General Equilibrium and Recession Phenomenon

**Nicholas S. Gonchar\*, Wolodymyr H. Kozyrski, Anatol S. Zhokhin**
Mathematical Modeling Department, The N. N. Bogolubov Institute for Theoretical Physics,
Ukrainian National Academy of Sciences, Kiev, Ukraine
\* Corresponding author E-mail address: mhonchar@i.ua  (N. S. Gonchar)

Abstract
The theorems we proved describe the structure of economic equilibrium in the exchange economy model. We have studied the structure of property vectors under given structure of demand vectors at which given price vector is equilibrium one. On this ground, we describe the general structure of the equilibrium state and give characteristic of equilibrium state describing economic recession. The theory developed is applied to explain the state of the economy in some European countries.

## 1. Introduction

There are many factors generating economic recessions. A knowledge of the various causes leading to unwanted economic evolution is very important. Skillful ability to confront these challenges is of essential priority.  For example, the monetary policy stimulating production growth is always accompanied by inflation.  So, taming it, the central bank can sometimes lead to economic fall. Another example is an artificial policy of exchange rate retention that may eventually lead to a sharp depreciation of the national currency.

We know many such economic destabilizing factors that are in some detail described by W. Semmler [1] where macroeconomic models are also exposed characterizing these phenomena. If these models can be a ground for the development of economic policy, they deserve attention. How can one at an early stage identify hidden factors of unwanted economic development? At the macroeconomic model level, it is hardly possible.  However, there is  a possibility of less aggregated description allowing to identify the key factors destabilizing economy.

*The subject of paper:* In this paper and in the previous ones [2,3,4], at microeconomic level under very general assumptions, we have established equilibrium state quality. *The scope of the subject:* When demand for a decisive group of goods is strictly less than the supply, the money partially loses its real value and the national currency devalues. And it does not depend on the nature of the factors that gave rise to it. It can be both ineffective monetary policy stimulating economic growth or the policy of raw materials export that led to the collapse of the economy after the fall of the national currency.  The main is to identify these trends at an early stage.

*The main goal of the paper* is to present here the theory we have developed to characterize economic equilibrium states describing the recession phenomenon.

*The novelty of research work* lies in its model grounds containing a new theory of information economics describing general equilibrium and characterizing economic recession.

*Research Significance.* The paper we present here has both theoretical and practical significance due to main Theorems underlying the theory of general equilibrium we developed and its predictive power to determine possible economic trend to recession state. The theory we give here can serve as an essential tool to study phenomena of economy decreasing.

We organized the paper as follows. In the Introduction, we give a short sketch of the problem history. In what follows, we introduce the notion of equilibrium quality. Then we state the problem and prove the main Theorems of the theory we developed. Then we present aggregated economy description in the general equilibrium. Next, we apply the theoretic model exposed to analyze the economic situation in some European countries. At rest, we end the paper with Conclusions and References.

Burns and Mitchell [5] identified a recession as a phenomenon when a wide range of economic indicators fell for a certain period, for example, for six months. The first phenomenological model containing cyclic fluctuations in the national product was proposed by P. A. Samuelson [6]. The theory of real business cycles (RBC) is a class of macroeconomic models in which fluctuations in the business cycle can be largely due to the real technology shocks. In contrast to the leading theories of business cycle theories, RBC theory understands the recession and periods of economic growth as an effective response to external changes in the real economic environment.

Let us cite here the Summers' statement [7] on the real business cycle models: "in my opinion, the real business cycle model by Kidland and Prescott [8] does not explain the phenomenon of cyclical economic development observed in the United States and other capitalist economies in the world."

The reason for it is as described in [7] that:

   1) Kydland and Prescott use the wrong parameters (one-third of time households devote to business activity, not one-sixth; historical real interest rates have been 4%, not 1%).

   2) The lack of independent evidence that technology shocks cause business cycles, and largely the impossibility to specify the technological reasons for the observed recessions.

   3) Kydland and Prescott model ignores commodity prices and the prediction of the asset values.

   4) Kydland and Prescott model ignores the exchange mechanism destruction (the inability of factories to sell their goods in exchange for labor).

RBC theory is the main cause of conflict between the macro-economists [5] because it categorically rejects Keynesian economics and the real efficiency of monetarist economy .

At all times, the recession was accompanied by the exchange mechanism destruction.

For the first time such a hypothesis was formulated in [5]: "It seems clear that a central aspect of depression and perhaps, more generally, economic fluctuations, is an exchange mechanism breaking.

Read any living account of the Great Depression time in the United States. The firms produced the goods and wanted to sell them. Workers were ready to sell labor in exchange for goods. However, the exchange did not happen."

How can one mathematically explain the destruction of the exchange mechanism? For this, we'll use a mathematical model of economic equilibrium, proposed and studied in [2,3] and examined in more detail in [4], where an assumption was first made on the mathematical aspect of the recession mechanism.

In the first part of the paper, we'll clarify the mathematical nature of the exchange mechanism destruction and prove the theorems confirming the hypotheses formulated in [3]. In the second part, we'll build a mathematical model of economic equilibrium describing adequately the economic equilibrium of the state economy. In the third part, we'll apply it to reveal recession within some European countries involving Ukraine.

## 2. About a quality of equilibrium state

Let us consider an exchange model with $l$ consumers. Each $i$-th consumer has some non-zero set of goods $b_i = \{b_{ki}\}_{k=1}^n$, $b_{ki} \geq 0$, $k = \overline{1,n}$. There are $n$ types of goods in the economic system. If some component $j$ of the vector $b_i$ vanishes, then the $i$-th consumer has not $j$-th type of goods. If the $i$-th consumer is individual, he/she has such a resource as a labor force.

Among consumers, there are firms that offer a set of products they produce and not only. For the economic system to operate, commodity exchange between consumers is needed. For firms, to buy resources from individuals to produce goods. For individuals, to buy goods for their needs.

Suppose that the $i$-th consumer having commodity set $b_i$ wants to exchange it for some commodity set proportional to the vector $C_i = \{C_{ki}\}_{k=1}^n$. We have studied such a model in [2,3,4] where we gave the necessary and sufficient conditions for equilibrium existence and constructed algorithms to find such states.

In this paper, we continue to study the model with the aim to explain and detect the recession. We assume that total supply in the model is $\psi = \{\psi_k\}_{k=1}^n$, $\psi_k = \sum_{i=1}^l b_{ki} > 0$, $k = \overline{1,n}$, where the first component of every vector $b_i$ is money supply and the first component of the vector $C_i$ is money demand.

**Definition 1.** We say that economic system in exchange model is in an economic equilibrium state [2] if there is such non-zero non-negative vector $p_0$ that the inequalities

$$\sum_{i=1}^l C_{ki} \frac{<b_i, p_0>}{<C_i, p_0>} \leq \psi_k, \quad k = \overline{1,n}, \qquad (1)$$

hold, where $<b_i, p_0> = \sum_{s=1}^n b_{si} p_s^0$, $<C_i, p_0> = \sum_{s=1}^n C_{si} p_s^0$.

Suppose the set of non-negative vectors $C_i = \{C_{ki}\}_{k=1}^n, i = \overline{1,l}$, satisfies the condition: there exists a such non-empty subset $I$ of the set $N = \{1, 2, ...n\}$ that $\sum_{k \in I} C_{ki} > 0, i = \overline{1,l}$. The problem is to describe a set of non-negative vectors $b_i, i = \overline{1,l}$, for which a non-negative vector $p_0 = \{p_i^0\}_{i=1}^n$ solves the set of equations

$$\sum_{i=1}^l C_{ki} \frac{<b_i, p_0>}{<C_i, p_0>} = \psi_k, \qquad k = \overline{1,n}. \tag{2}$$

if components $p_i^0, i \in I$, of the vector $p_0 = \{p_i^0\}_{i=1}^n$ are strictly positive ones and other components $p_i^0 = 0, i \in N \setminus I = J$, and the set of non-negative vectors $C_i = \{C_{ki}\}_{k=1}^n, i = \overline{1,l}$, satisfy the conditions above.

**Theorem 1.** Assume the set of non-negative vectors $C_i = \{C_{ki}\}_{k=1}^n, i = \overline{1,l}$, satisfies the conditions $\sum_{k \in I} C_{ki} > 0, i = \overline{1,l}$, where $I$ is non-empty subset of $N$ and the vector $\psi = \sum_{i=1}^l b_i = \{\psi_k\}_{k=1}^n, \psi_k > 0, k \in I$, belongs to a positive cone formed by vectors $C_i = \{C_{ki}\}_{k=1}^n, i = \overline{1,l}$. The necessary and sufficient condition that the vector $p_0 = \{p_i^0\}_{i=1}^n, p_i^0 > 0, i \in I, p_i^0 = 0, i \in J = N \setminus I$ solves the set of equations (2) is the representation for the set of vectors $b_i, i = \overline{1,l}$,

$$b_i = y_i \frac{<C_i, p_0>}{<\psi, p_0>} \psi + \sum_{s \in I} a_s^i (e_s - \frac{p_s^0}{\sum_{s \in I} p_s^0} e_I) + d_i^0, \quad i = \overline{1,l}, \tag{3}$$

where the vector $y = \{y_i\}_{i=1}^l, y_i \geq 0$, solves the set of equations

$$\sum_{i=1}^l C_i y_i = \psi, \tag{4}$$

$e_i = \{\delta_{ij}\}_{j=1}^n, \delta_{ij}$ is the Kronecker symbol, $e_I = \sum_{i \in I} e_i, \sum_{i=1}^l a_s^i = 1, s \in I$, and the set of vectors $d_i^0 = \{d_{ki}^0\}_{k=1}^n, i = \overline{1,l}$, satisfies the conditions $d_{ki}^0 = 0, k \in I, i = \overline{1,l}, \sum_{i=1}^l d_i^0 = 0$.

**Proof. Necessity.** Let us suppose that $p_0$ solves the set of equations (2) and satisfies Theorem 1 conditions. Let

$$y_i = \frac{<b_i, p_0>}{<C_i, p_0>}, \quad i = \overline{1,l}. \tag{5}$$

The vector $y = \{y_i\}_{i=1}^l, y_i \geq 0$, satisfies the set of equations (4). Introduce the notation

$$d_i = b_i - y_i \frac{<C_i, p_0>}{<\psi, p_0>} \psi, \quad i = \overline{1,l}. \tag{6}$$

Then

$$<d_i, p_0> = 0, \quad i = \overline{1,l}, \quad \sum_{i=1}^{l} d_i = 0. \qquad (7)$$

The vector set (6) satisfying conditions (7) can be presented as the sum of two vectors $d_i^1 + d_i^0$ where the vector $d_i^1 = \{d_{ki}^1\}_{k=1}^n$ is such that $d_{ki}^1 = 0$, $k \in J$, and components of the vector $d_i^0 = \{d_{ki}^0\}_{k=1}^n$ satisfy the conditions $d_{ki}^0 = 0$, $k \in I$, $i = \overline{1,l}$. To satisfy the conditions (7), we have to require for the set of vectors $d_i^1$ carrying out of conditions

$$<d_i^1, p_0> = 0, \quad i = \overline{1,l}, \quad \sum_{i=1}^{l} d_i^1 = 0, \qquad (8)$$

and for the set of vectors $d_i^0$ carrying out of the condition

$$\sum_{i=1}^{l} d_i^0 = 0, \qquad (9)$$

since the conditions $<d_i^0, p_0> = 0$, $i = \overline{1,l}$, are valid. Describe all the vectors satisfying conditions (8). Consider the set of vectors

$$g_s = e_s - \frac{p_s^0}{\sum_{s \in I} p_s^0} e_I, \quad s \in I. \qquad (10)$$

The vectors $g_s$, $s \in I$, satisfy the conditions

$$<g_s, p_0> = 0, \quad s \in I, \quad \sum_{s \in I} g_s = 0. \qquad (11)$$

It is easy to show that the set of vectors $g_s$, $s \in I$, has the rank $|I|-1$, where $|I|$ is the number of elements in the set $I$. Therefore, for every vector $d_i^1$ there is unique representation

$$d_i^1 = \sum_{s_j=1}^{|I|-1} h_{s_j}^i g_{s_j}. \qquad (12)$$

From the condition (8), we have

$$0 = \sum_{i=1}^{l} d_i^1 = \sum_{s_j=1}^{|I|-1} \sum_{i=1}^{l} h_{s_j}^i g_{s_j}. \qquad (13)$$

As a consequence of linear independence of the vectors $g_{s_1}, g_{s_2}, \ldots g_{s_{|I|-1}}$, we obtain $\sum_{i=1}^{l} h_{s_j}^i = 0$, $s_j = \overline{1, |I|-1}$. As $\sum_{s_j \in I} g_{s_j} = 0$, then, adding to (12) the vector $\frac{1}{l} \sum_{s_j \in I} g_{s_j} = 0$, we have

$$d_i^1 = \sum_{s_j=1}^{|I|-1} (h_{s_j}^i + \frac{1}{l}) g_{s_j} + \frac{1}{l} g_{s_{|I|}} = \sum_{s \in I} a_s^i g_s.$$

Then it is obvious that $\sum_{i=1}^{l} a_s^i = 1$. By this, we proved the representation for the vector $d_i^1$. It is evident that $\sum_{i=1}^{l} d_i^1 = 0$. The necessity is proven.

**Sufficiency.** Assume the representation (3) holds. Then $<b_i, p_0> = y_i <C_i, p_0>$. From the Theorem conditions $<C_i, p_0> \neq 0$. Therefore, $y_i = \dfrac{<b_i, p_0>}{<C_i, p_0>}$. Substituting $y_i$ into (4) we have needed statement.

Theorem 1 is proven.

Theorem 1 is the Theorem about market clearing. Having fixed demand structure determined by the vector set $C_i = \{C_{ki}\}_{k=1}^n, i = \overline{1,l}$, we have found the necessary and sufficient conditions for the supply $b_i = \{b_{ki}\}_{k=1}^n, i = \overline{1,l}$, structure under which the demand is equal to the supply if the equilibrium price vector is known. The Theorem is important as an instrument allowing for the fixed demand structure to find out the supply structure that gives equilibrium state degeneracy explaining recession phenomenon.

**Definition 2.** At given equilibrium price vector $p_0$, property distribution $b_i$, $i = \overline{1,l}$, in society is equivalent to that $\bar{b}_i$, $i = \overline{1,l}$, if there exists a set of vectors $d_i$, $i = \overline{1,l}$, satisfying conditions $<d_i, p_0> = 0$, $i = \overline{1,l}$, and for the vectors $\bar{b}_i$, $i = \overline{1,l}$, the representation

$$\bar{b}_i = b_i + d_i, \quad i = \overline{1,l}.$$

is valid.

Let us note that at equilibrium equivalent property distributions have the same value. Therefore, the Theorem 2 holds.

**Theorem 2.** If the vector $p_0$ is an equilibrium price vector satisfying the set of equations (2), then it is also an equilibrium price vector for the equivalent property distribution $\bar{b}_i$, $i = \overline{1,l}$, and satisfies the same set of equations (2) with property vectors $\bar{b}_i$, $i = \overline{1,l}$, under condition that $\sum_{i=1}^{l} d_i = 0$.

**Theorem 3.** Suppose that the Theorem 1 conditions hold and $p_0$ has positive components with indices belonging to the set $I$ and solves the set of equations (2). Then a set of such vectors $\tilde{d}_i^0$, $i = \overline{1,l}$, exists that $<\tilde{d}_i^0, p_0> = 0$, $i = \overline{1,l}$, $\sum_{i=1}^{l} \tilde{d}_i^0 = 0$, and an equivalent property distribution $\bar{b}_i = b_i + \tilde{d}_i^0$, $i = \overline{1,l}$, is such that the rank of the vector set $\bar{b}_i - y_i C_i$, $i = \overline{1,l}$, does not exceed $|I|$. The set of equations

$$\sum_{i=1}^{l} C_{ki} \frac{<\bar{b}_i, p_0>}{<C_i, p_0>} = \sum_{i=1}^{l} \bar{b}_{ki}, \qquad k = \overline{1,n},$$

has branching solution $p_0$ whose degeneracy multiplicity is not less than $n - |I|$, where $\bar{b}_i = \{\bar{b}_{ki}\}_{k=1}^n$,

$i = \overline{1,l}$. The value of the goods whose indices belong to the set $J$ can be arbitrary one.

**Proof.** Construct the vector set $\tilde{d}_i^0$, $i = \overline{1,l}$, declared in the Theorem 3. Introduce the vector $\psi_0 = \{\psi_k^0\}_{k=1}^n$ where $\psi_k^0 = 0$, $k \in I$, $\psi_k^0 = \psi_k$, $k \in J$, and the set of vectors $C_i^0 = \{C_{ki}^0\}_{k=1}^n$, $C_{ki}^0 = 0$, $k \in I$, $C_{ki}^0 = C_{ki}$, $k \in J$, $i = \overline{1,l}$. Suppose

$$\overline{d}_i^0 = -y_i \frac{<C_i, p_0>}{<\psi, p_0>}\psi_0 + y_i C_i^0, \quad i = \overline{1,l}.$$

It is evident that $<\overline{d}_i^0, p_0> = 0$, $i = \overline{1,l}$, $\sum_{i=1}^l \overline{d}_i^0 = 0$. Moreover, $\overline{d}_i^0 = \{\overline{d}_{ki}^0\}_{k=1}^n$, $\overline{d}_{ki}^0 = 0$, $k \in I$. From the Theorem 1, the validity of the representation (3) follows for the vectors $b_i$, $i = \overline{1,l}$. The set of vectors $\overline{d}_i^0$, $i = \overline{1,l}$, we have constructed is such as the vector set $d_i^0$, $i = \overline{1,l}$, in the Theorem 1. Therefore, introduce the vector $\overline{b}_i = b_i + \tilde{d}_i^0$, $\tilde{d}_i^0 = \overline{d}_i^0 - d_i^0$ where the vector $d_i^0$ enters vector $b_i$ representation. Then, having representation (3), we obtain

$$\overline{b}_i = y_i \frac{<C_i, p_0>}{<\psi, p_0>}\psi + \sum_{s \in I} a_s^i (e_s - \frac{p_s^0}{\sum_{s \in I} p_s^0} e_I) - y_i \frac{<C_i, p_0>}{<\psi, p_0>}\psi_0 + y_i C_i^0, \quad i = \overline{1,l}.$$

Obviously, $\overline{b}_i = \{\overline{b}_{ki}\}_{k=1}^n \geq 0$, $i = \overline{1,l}$. Really, the vectors $\overline{d}_i^0 - d_i^0$, $i = \overline{1,l}$, have zero components at the set of indices $I$, therefore, $\overline{b}_{ki} = b_{ki} \geq 0$, $k \in I$. If $k \in J$, then $\overline{b}_{ki} = y_i C_{ki}$. In fact, as $\psi_k^0 = \psi_k$, $C_{ki}^0 = C_{ki}$, $k \in J$, $i = \overline{1,l}$, then components of the vectors

$$\sum_{s \in I} a_s^i (e_s - \frac{p_s^0}{\sum_{s \in I} p_s^0} e_I), \quad y_i \frac{<C_i, p_0>}{<\psi, p_0>}\psi - y_i \frac{<C_i, p_0>}{<\psi, p_0>}\psi_0$$

vanish at the set of indices $J$. Therefore, for such price vector $p = \{p_k\}_{k=1}^n$ that $p_k = p_k^0$, $k \in I$, and components $p_k$, $k \in J$, being arbitrary non-negative numbers, we have

$$<\overline{b}_i, p> = \sum_{k \in I} \overline{b}_{ki} p_k^0 + \sum_{k \in J} \overline{b}_{ki} p_k = <b_i, p_0> + \sum_{k \in J} y_i C_{ki} p_k = y_i <C_i, p_0> + y_i \sum_{k \in J} C_{ki} p_k =$$

$$y_i [\sum_{k \in I} C_{ki} p_k^0 + \sum_{k \in J} C_{ki} p_k] = y_i <C_i, p>.$$

Therefore, $y_i = \frac{<\overline{b}_i, p>}{<C_i, p>}$. Further, as $\sum_{i=1}^l \overline{d}_i^0 = 0$, $\sum_{i=1}^l d_i^0 = 0$, then $\sum_{i=1}^l \overline{b}_i = \sum_{i=1}^l b_i$.

From here, we obtain that the vector $p$ solves the set of equations

$$\sum_{i=1}^l C_{ki} \frac{<\overline{b}_i, p>}{<C_i, p>} = \sum_{i=1}^l b_{ki}, \quad k = \overline{1,n}.$$

From equalities $<\overline{b}_i - y_i C_i, p> = 0$, $i = \overline{1,l}$, it follows that the number of linearly independent solutions of the last set of equations is not less than $n-|I|$ that means that a degeneracy multiplicity of equilibrium state is not less than $n-|I|$.

The Theorem 3 is proven.

In this Theorem, basing on the above introduced notion of equivalent property distribution, we show the existence of such equivalent property distribution in society for which multiple degeneracy of economic equilibrium happens. In such a case, random factors can provoke the transition between any possible equilibrium states.

The Theorem proven confirms our assumption from [3] that at the equilibrium state the recession is accompanied by equilibrium state degeneracy. From the Theorem 3, it follows that at equilibrium point $p_0$ there is branching of solutions, i.e., there is $r = n-|I|$ -parametric family of solutions being family of equilibrium states. In this case, let us introduce the notion of real value for the national currency unit to characterize such equilibrium states.

In the model considered, we suppose that the first component $p_1^0$ of the equilibrium price vector $p_0$ in the economic system is the nominal value of such specific goods as money, which we take equal one as equilibrium price vector is determined up to positive factor and the first component $\psi_1$ of the supply vector $\psi$ is the supply of money in the economic system.

Let us determine the real money value for the equilibrium price vector $p_0$ supposing

$$\overline{p}_0^{-1} = \frac{\sum_{i=2}^{n} p_i^0 \psi_i}{\psi_1}.$$

If the degeneracy multiplicity of equilibrium state $p_0$ equals $r = 1,$ then the real money value is determined uniquely. In that case, money will be both a medium of exchange and a means to save the value. If $r > 1$, then given property distribution in economic system corresponds to a family of equilibrium states.

In this case, money has eroded value because the last formula gives a family of values. If the fluctuation of the real money value is insignificant at given $r$, then money has both exchange function and approximate value function too. In the opposite case, i.e., when criticality becomes such that money loses in part its value function, national currency devalues and, as we see, the reason is a discrepancy between supply and demand structures in the economic system. In such economic system, reform is needed for the property distribution structure, i.e., structural economic transformation.

We give here the Theorem from [3] being necessary to study the general equilibrium structure.

**Theorem 4.** Let the conditions $\sum_{k=1}^{n} C_{ki} > 0, \ i = \overline{1,l}, \ \sum_{i=1}^{l} C_{ki} > 0, \ i = \overline{1,n}$ hold. The necessary and sufficient conditions for an equilibrium to exist in the exchange model are the following ones: there exist such non-zero non-negative vector $y = \{y_i\}_{i=1}^l$ and non-zero non-negative vector $\bar{\psi} = \{\bar{\psi}_k\}_{k=1}^n$ that

$$\bar{\psi} = \sum_{i=1}^l y_i C_i, \tag{14}$$

and for the vectors $b_i$ the following representation holds

$$b_i = \bar{b}_i + d_i, \quad \bar{b}_i = y_i \frac{<C_i, p_0>}{<\bar{\psi}, p_0>}\psi, \quad i = \overline{1,l}, \quad \sum_{i=1}^l d_i = 0, \quad <p_0, d_i> = 0, \quad i = \overline{1,l}, \tag{15}$$

where $p_0$ is non-zero non-negative vector satisfying the conditions

$$<\bar{\psi}, p_0> = <\psi, p_0>, \quad <C_i, p_0> > 0, \quad i = \overline{1,l}, \quad \bar{\psi} \leq \psi. \tag{16}$$

**Theorem 5.** Suppose that Theorem 4 conditions hold and $p_0 = \{p_i^0\}_{i=1}^n$ is such equilibrium price vector that

$$\sum_{i=1}^l C_{ki} \frac{<b_i, p_0>}{<C_i, p_0>} = \psi_k, \quad k \in I, \tag{17}$$

$$\sum_{i=1}^l C_{ki} \frac{<b_i, p_0>}{<C_i, p_0>} < \psi_k, \quad k \in J. \tag{18}$$

Then $p_i^0 = 0$, $i \in J$, where $N = I \cup J$, $I \cap J = \Theta$, $\Theta$ – empty set and $I$ – non-empty subset of $N$.

See Proof of the Theorem in [2].

**Theorem 6.** Assume that the Theorem 4 conditions hold and $p_0$ having positive components with the indices belonging to the set $I$ solves the inequality set (1). Then there is vector set, $\tilde{d}_i^0$, $i = \overline{1,l}$, such that $<\tilde{d}_i^0, p_0> = 0$, $i = \overline{1,l}$, $\sum_{i=1}^l \tilde{d}_i^0 \leq 0$, and an equivalent property distribution $\bar{b}_i = b_i + \tilde{d}_i^0$, $i = \overline{1,l}$, is such that the rank of the vector set $\bar{b}_i - y_i C_i$, $i = \overline{1,l}$, does not exceed $|I|$, and $\sum_{i=1}^l \bar{b}_i = \psi$. The set of equations

$$\sum_{i=1}^l C_{ki} \frac{<\bar{b}_i, p_0>}{<C_i, p_0>} = \bar{\psi}_k, \quad k = \overline{1,n},$$

has a branching solution $p_0$ whose degeneracy multiplicity is not less than $n - |I|$, where $\bar{b}_i = \{\bar{b}_{ki}\}_{k=1}^n$, $i = \overline{1,l}$. The value of goods whose indices belong to the set $J$ can be arbitrary one.

**Proof.** By the Theorem 4, the vector $p_0$ solves the set of equations

$$\sum_{i=1}^l C_{ki} \frac{<b_i, p_0>}{<C_i, p_0>} = \psi_k, \quad k = \overline{1,n}.$$

Denote

$$y_i = \frac{<b_i, p_0>}{<C_i, p_0>}, \quad i = \overline{1,l}.$$

The vector $y = \{y_i\}_{i=1}^{l}$, $y_i \geq 0$, solves the set of equations (14). Introduce the notation

$$d_i = b_i - y_i \frac{<C_i, p_0>}{<\overline{\psi}, p_0>} \overline{\psi}, \quad i = \overline{1,l}. \qquad (19)$$

Then

$$<d_i, p_0> = 0, \quad i = \overline{1,l}, \quad \sum_{i=1}^{l} d_i \geq 0. \qquad (20)$$

The set of vectors (19) satisfying conditions (20) can be expressed as the sum of two vectors $d_i^1 + d_i^0$, where the vector $d_i^1 = \{d_{ki}^1\}_{k=1}^{n}$ is such that $d_{ki}^1 = 0$, $k \in J$, and components of the vector $d_i^0 = \{d_{ki}^0\}_{k=1}^{0}$ satisfy the conditions $d_{ki}^0 = 0$, $k \in I$, $i = \overline{1,l}$. To satisfy conditions (20), we need to require that the set of vectors $d_i^1$, $d_i^0$ satisfied the conditions

$$<d_i^1, p_0> = 0, \quad i = \overline{1,l}, \quad \sum_{i=1}^{l} d_i^1 = 0,$$

$\sum_{i=1}^{l} d_i^0 \geq 0$, as the conditions $<d_i^0, p_0> = 0$, $i = \overline{1,l}$, are valid.

As in the Theorem 1 Proof, from this we obtain the representation for the vector set $b_i$, $i = \overline{1,l}$,

$$b_i = y_i \frac{<C_i, p_0>}{<\overline{\psi}, p_0>} \overline{\psi} + \sum_{s \in I} a_s^i (e_s - \frac{p_s^0}{\sum_{s \in I} p_s^0} e_I) + d_i^0, \quad i = \overline{1,l},$$

where vector set $d_i^0$, $i = \overline{1,l}$, satisfies the conditions $\sum_{i=1}^{l} d_i^0 \geq 0$, $<d_i^0, p_0> = 0$, $i = \overline{1,l}$.

Let us construct the vector set $\tilde{d}_i^0$, $i = \overline{1,l}$, declared in the Theorem 5. Introduce a vector $\overline{\psi}_0 = \{\overline{\psi}_k^0\}_{k=1}^{n}$, where $\overline{\psi}_k^0 = 0$, $k \in I$, $\overline{\psi}_k^0 = \overline{\psi}_k$, $k \in J$, and a vector family $C_i^0 = \{C_{ki}^0\}_{k=1}^{n}$, $C_{ki}^0 = 0$, $k \in I$, $C_{ki}^0 = C_{ki}$, $k \in J$, $i = \overline{1,l}$. Let us put

$$\overline{d}_i^0 = -y_i \frac{<C_i, p_0>}{<\overline{\psi}, p_0>} \overline{\psi}_0 + y_i C_i^0, \quad i = \overline{1,l}.$$

It is evident that $<\overline{d}_i^0, p_0> = 0$, $i = \overline{1,l}$, $\sum_{i=1}^{l} \overline{d}_i^0 = 0$. Moreover, $\overline{d}_i^0 = \{\overline{d}_{ki}^0\}_{k=1}^{n}$, $\overline{d}_{ki}^0 = 0$, $k \in I$. Introduce the vector $\overline{b}_i = b_i + \tilde{d}_i^0$, $\tilde{d}_i^0 = \overline{d}_i^0 - d_i^0$ where the vector $d_i^0$ enters the representation for the vector $b_i$. Then we obtain

$$\overline{b}_i = y_i \frac{<C_i, p_0>}{<\overline{\psi}, p_0>} \overline{\psi} + \sum_{s \in I} a_s^i (e_s - \frac{p_s^0}{\sum_{s \in I} p_s^0} e_I) - y_i \frac{<C_i, p_0>}{<\overline{\psi}, p_0>} \overline{\psi}_0 + y_i C_i^0, \quad i = \overline{1,l}.$$

It is obvious that $\overline{b}_i = \{\overline{b}_{ki}\}_{k=1}^{n} \geq 0$, $i = \overline{1,l}$. It follows from that the vectors $\overline{d}_i^0 - d_i^0$, $i = \overline{1,l}$, have zero components at the set of indices $I$. Therefore, $\bar{b}_{ki} = b_{ki} \geq 0$, $k \in I$. If $k \in J$, then $\bar{b}_{ki} = y_i C_{ki}$. In fact, as $\bar{\psi}_k^0 = \bar{\psi}_k$, $C_{ki}^0 = C_{ki}$, $k \in J$, $i = \overline{1,l}$, then components of the vectors

$$\sum_{s \in I} a_s^i (e_s - \frac{p_s^0}{\sum_{s \in I} p_s^0} e_l), \quad y_i \frac{<C_i, p_0>}{<\overline{\psi}, p_0>} \overline{\psi} - y_i \frac{<C_i, p_0>}{<\overline{\psi}, p_0>} \overline{\psi}_0$$

vanish at the set of indices $J$. Therefore, for such price vector $p = \{p_k\}_{k=1}^n$ that $p_k = p_k^0$, $k \in I$, and components $p_k$, $k \in J$, are arbitrary non-negative numbers, we have

$$<\bar{b}_i, p> = \sum_{k \in I} \bar{b}_{ki} p_k^0 + \sum_{k \in J} \bar{b}_{ki} p_k = <b_i, p_0> + \sum_{k \in J} y_i C_{ki} p_k = y_i <C_i, p_0> + y_i \sum_{k \in J} C_{ki} p_k =$$

$$y_i [\sum_{k \in I} C_{ki} p_k^0 + \sum_{k \in J} C_{ki} p_k] = y_i <C_i, p>.$$

Therefore, $y_i = \dfrac{<\bar{b}_i, p>}{<C_i, p>}$. Further, as $\sum_{i=1}^l \bar{d}_i^0 = 0$, $\sum_{i=1}^l d_i^0 = 0$, then $\sum_{i=1}^l \bar{b}_i = \sum_{i=1}^l b_i$.

From here, we obtain that the vector $p$ solves the set of equations

$$\sum_{i=1}^l C_{ki} \frac{<\bar{b}_i, p>}{<C_i, p>} = \overline{\psi}_k, \quad k = \overline{1,n}.$$

From equalities $<\bar{b}_i - y_i C_i, p> = 0$, $i = \overline{1,l}$, it follows that the number of linearly independent solutions to the last equation set is not less than $n - |I|$ having mean that the degeneracy multiplicity of equilibrium state is not less than $n - |I|$.

Theorem 6 is proven.

The Theorem 6 meaning is that at equilibrium state there exists such equivalent property distribution under which the demand for goods whose indices belong to the set $J$ is the same as their supply, i.e., components of the vectors $\bar{b}_i$ and $y_i C_i$ for the indices from $J$ coincide. From this, it follows that the value of such goods is not determined by equilibrium condition. And, as we note above, in this case money loses in part its functions of exchange and value. Degeneracy multiplicity of equilibrium state in this case is not less than $|J|$.

If this state significantly destabilizes the economy, then devaluation of the national currency happens inevitably and all the problems related occur, namely, increasing unemployment and devaluation of deposits. From the quality of this equilibrium state, it follows that a further increase in money supply, even at a fairly low interest rate without changing the structure of investment will not lead to economic growth. What is needed is a cardinal change of economic structure, investments into new perspective industries, and creation of new jobs in these sectors. Therefore, recession state is such equilibrium state when a significant part of goods produced are not selling which in turn leads to the decline of many indicators of the economic state. The quality of this equilibrium is such that there is a breakdown of the exchange mechanism.

In the next two Theorems, we give sufficient conditions for the existence of such equilibrium at which demand equals supply.

**Theorem 7.** Let the matrix $B = |b_{ki}|_{k=1,i=1}^{n,l}$, whose columns are the vectors $b_i = \{b_{ki}\}_{k=1}^n$, $i = \overline{1,l}$, be expressed as $B = CB_1$, where the matrix $B_1 = |b_{ki}^1|_{k,i=1}^l$ is non-negative and indecomposable one and the matrix $C = |C_{ki}|_{k=1,i=1}^{n,l}$ is composed of columns $C_i = \{C_{ki}\}_{k=1}^n$, $i = \overline{1,l}$, and is such that $\sum_{k=1}^n C_{ki} > 0$, $i = \overline{1,l}$.

Then there exists a strictly positive solution to the problem

$$\sum_{k=1}^l b_{ks}^1 d_k = y_s d_s, \quad s = \overline{1,l}, \tag{21}$$

with respect to the vector $d = \{d_k\}_{k=1}^l$, where $y_k = \sum_{s=1}^l b_{ks}^1$. If the vector $d$ belongs to the interior of the cone generated by the vectors $C_i^T = \{C_{ik}\}_{k=1}^l$, $i = \overline{1,n}$, then there exists a problem (2) solution solving the problem

$$\sum_{k=1}^l C_{ki} p_k = d_i, \quad i = \overline{1,l}. \tag{22}$$

**Proof.** The problem conjugate to the problem (21)

$$\sum_{s=1}^l b_{ks}^1 r_s = y_k r_k, \quad k = \overline{1,l}, \tag{23}$$

has a solution $r = \{1,...,1\} \in R^l$. As the matrix $B_1$ is non-negative and indecomposable one, the problem

$$\sum_{s=1}^l \frac{b_{ks}^1}{y_k} r_s = r_k, \quad k = \overline{1,l}, \tag{24}$$

has unique up to a constant factor solution. Therefore, by the Perron-Frobenius theorem, there is a strictly positive solution to the conjugate problem

$$\sum_{k=1}^l \frac{b_{ks}^1}{y_k} \overline{d}_k = \overline{d}_s, \quad s = \overline{1,l}.$$

Let us put $d = \left\{\dfrac{\overline{d}_k}{y_k}\right\}_{k=1}^l$. Then the vector $d$ is a strictly positive solution to the problem (21). By the Theorem assumptions, there is a strictly positive solution $p_0 = \{p_i^0\}_{i=1}^n$ to the problem (22). Substituting vector $d$ into (21) and taking into account (22), we obtain that the vector $p_0 = \{p_i^0\}_{i=1}^n$ solves the problem

$$\sum_{i=1}^n (b_{is} - y_s C_{is}) p_i^0 = 0, \quad s = \overline{1,l}.$$

The last means that the vector $p_0 = \{p_i^0\}_{i=1}^n$ is a strictly positive solution to the set of equations (2).

**Theorem 8.** Let the matrix $C = |C_{ki}|_{k=1,i=1}^{n,l}$ be composed of columns $C_i = \{C_{ki}\}_{k=1}^n$, $i = \overline{1,l}$, and such that $\sum_{k=1}^n C_{ki} > 0$, $i = \overline{1,l}$, and the matrix $B = |b_{ki}|_{k=1,i=1}^{n,l}$, whose columns are vectors $b_i = \{b_{ki}\}_{k=1}^n$, $i = \overline{1,l}$, be expressed as $B = CB_1$, where the matrix $B_1 = |b_{ki}^1|_{k,i=1}^l$ is such that $y_k = \sum_{s=1}^l b_{ks}^1 \geq 0$, $\overline{y}_j = \sum_{k=1}^l b_{kj}^1 \geq 0$, $j,k = \overline{1,l}$. If the vector $\overline{y} = \{\overline{y}_j\}_{j=1}^l$ solves the problem

$$\sum_{i=1}^l C_i \overline{y}_i = \psi, \qquad (25)$$

and the vector $e = \{1,1,...,1\} \in R^l$ belongs to the interior of the cone generated by the vectors $C_i^T = \{C_{ik}\}_{k=1}^l$, $i = \overline{1,n}$, then there exists a strictly positive equilibrium vector $p_0 = \{p_i^0\}_{i=1}^n$ solving the problem (2).

**Proof.** As

$$<b_j, p> = \sum_{i=1}^l <C_i, p> b_{ij}^1,$$

we'll demand the validity of equality $\dfrac{<b_j, p>}{<C_j, p>} = \overline{y}_j$, $j = \overline{1,l}$, or

$$\sum_{i=1}^l <C_i, p> b_{ij}^1 = <C_j, p> \overline{y}_j.$$

To satisfy the last equality, suppose

$$<C_i, p> = 1, \qquad i = \overline{1,l}. \qquad (26)$$

However, equation set (26) has a strictly positive solution by the Theorem assumption. Theorem 8 is proven.

**Corollary 1.** If the matrix $B_1$ in the Theorem 8 is, moreover, symmetric one, then the vector $\overline{y}$ appearing in the Theorem 8 solves the set of equations (25).

## 3. Economic Equilibrium within Aggregated Economy Description

Suppose that the economy, as earlier, produces $n$ types of goods and contains $l$ consumers. We'll say that the economy description is aggregated up to $m$ pure industries if the set $N = \{1,2,...,n\}$ is a union of such non-empty subsets $N_i$, $i = \overline{1,m}$, that $N = \bigcup_{i=1}^m N_i$, $N_i \cap N_j = \theta$, $i \neq j$, $\theta$ is empty set and a mapping $U$ of the set $R_+^n$ into the set $R_+^m$ is given by the rule: $x^u = Ux$, где $x = \{x_i\}_{i=1}^n$, $x^u = \{x_i^u\}_{i=1}^m$, $x_k^u = \sum_{s \in N_k} x_s$, $k = \overline{1,m}$.

If we characterize, as in the previous Section, $l$ consumers by property vectors $b_i = \{b_{ki}\}_{k=1}^n$ and demand vectors $C_i = \{C_{ki}\}_{k=1}^n$, $i = \overline{1,l}$, then within aggregated description every consumer will have aggregated characteristics, namely, property vectors $b_i^u = \{b_{ki}^u\}_{k=1}^m$ and demand vectors $C_i^u = \{C_{ki}^u\}_{k=1}^m$, $i = \overline{1,l}$. Let an economy be in an economic equilibrium state with the equilibrium price vector $p_0 = \{p_i^0\}_{i=1}^n$, then

$$\sum_{i=1}^l C_{ki} \frac{<b_i, p_0>}{<C_i, p_0>} \leq \psi_k, \quad k = \overline{1,n},$$

where $<b_i, p_0> = \sum_{s=1}^n b_{si} p_s^0$, $<C_i, p_0> = \sum_{s=1}^n C_{si} p_s^0$, $\psi = \{\psi_k\}_{k=1}^n$, $\psi_k = \sum_{i=1}^l b_{ki} > 0$, $k = \overline{1,n}$.

We can rewrite the last inequality set in the aggregated form

$$\sum_{i=1}^l C_{ki}^u \frac{<b_i^u, p_0>}{<C_i^u, p_0>} \leq \psi_k^u, \quad k = \overline{1,m}. \tag{27}$$

**Definition 3.** We say that an aggregation up to $m$ pure industries in economy is agreed with equilibrium state described in disaggregated way if there exists such aggregated equilibrium vector $p^u = \{p_i^u\}_{i=1}^m$ that

$$\sum_{i=1}^l C_{ki}^u \frac{<b_i^u, p^u>}{<C_i^u, p^u>} \leq \psi_k^u, \quad k = \overline{1,m}, \tag{28}$$

and, moreover, equalities in (27) and (28) hold for the same indices $k = \overline{1,m}$.

Further, we build mathematical model of economic equilibrium at the level of state. Suppose, the state's economy is described by $m$ pure industries each of which produces one type of goods. Production structure is described by the Leontief productive input-output matrix $A = |a_{ik}|_{i,k=1}^m$. Let the gross output vector in an economy is $x = \{x_i\}_{i=1}^m$, where $x_i$ is the gross output of the $i$-th pure industry.

Suppose that in an open economy the interindustry balance

$$x_k = \sum_{j=1}^m a_{kj} x_j + c_k^f + e_k - i_k, \quad k = \overline{1,m},$$

holds, where $c^f = \{c_k^f\}_{i=1}^m$ is a final consumption vector consisting of the sum of household final consumption vectors and the vector of gross capital formation and inventory changes, $e = \{e_k\}_{k=1}^m$ − export vector, $i = \{i_k\}_{k=1}^m$ is import vector. Let $p = \{p_i\}_{i=1}^m$ be a price vector, where $p_i$ is the price for unit of goods produced by the $i$-th industry. The pure $i$-th industry of the state's economy forms the demand for the resources of households determined by the vectors $C_i = \{x_i a_{ki}\}_{k=1}^m$, $i = \overline{1,m}$.

Supply vector of $i$-th industry is $b_i = \{\delta_{ik} x_k\}_{k=1}^m$, where $\delta_{ik}$ is the Kronecker symbol. The value of the gross product produced by $i$-th industry is $<b_i, p> = x_i p_i$, and new produced value by $i$-th industry equals $x_i(p_i - \sum_{s=1}^{m} a_{si} p_s)$, $i = \overline{1,m}$. To provide the production, households of the *i*-th industry form a resource supply $b_{m+i} = \{\delta_{ik} \sum_{j=1}^{m} a_{kj} x_j\}_{k=1}^{m}$, whose value is $<b_{m+i}, p> = p_i \sum_{j=1}^{m} a_{ij} x_j$, $i = \overline{1,m}$. If $\pi = \{\pi_i\}_{i=1}^{m}$ is a taxation vector [2], we suppose that only part of new produced value $\pi_i x_i (p_i - \sum_{s=1}^{m} a_{si} p_s)$ is used to produce final consumption goods, goods to extend the production and goods for export, meanwhile a part $(1-\pi_i) x_i (p_i - \sum_{s=1}^{m} a_{si} p_s)$ of value made by the *i*-th industry is used for social consumption, renewal of fixed assets, capitalization, infrastructure, public utilities and so on.

We suppose that at the market of goods of final consumption, households of the *i*-th industry form a demand proportional to value of sold resources and part of the industry deductions to consume the final consumption goods, the acquisition of fixed assets, capitalization, infrastructure, i. e.,

$$C_{m+i} = \left\{ \frac{[(1-\pi_i) x_i (p_i - \sum_{s=1}^{m} a_{si} p_s) + <b_{m+i}, p>] c_j^f}{\sum_{k=1}^{n} <b_{m+k}, p> + \sum_{i=1}^{m} x_i (p_i - \sum_{s=1}^{m} a_{si} p_s)} \right\}_{j=1}^{m}, \quad i = \overline{1,m}.$$

At rest, at the state's market, a foreign trade agent forms goods supply $b_{2m+1} = i = \{i_k\}_{k=1}^{m}$ – import vector, forming at the same time the demand for goods produced in the state $C_{2m+1} = e = \{e_k\}_{k=1}^{m}$ – export vector. Then the equilibrium price vector is determined by the condition that the demand does not exceed the supply $\sum_{i=1}^{m} x_i a_{ki} \frac{\pi_i x_i (p_i - \sum_{s=1}^{m} a_{si} p_s)}{\sum_{s=1}^{m} x_i a_{si} p_s} + c_k^f \frac{\sum_{i=1}^{m}(1-\pi_i) x_i (p_i - \sum_{s=1}^{m} a_{si} p_s) + \sum_{i=1}^{m} p_i \sum_{j=1}^{m} a_{ij} x_j}{\sum_{s=1}^{m} c_s^f p_s} + e_k \frac{\sum_{s=1}^{m} i_s p_s}{\sum_{s=1}^{m} e_s p_s} \leq$

$x_k + i_k,$

$$k = \overline{1,m},$$

or

$$\sum_{i=1}^{m} x_i a_{ki} \frac{\pi_i x_i p_i}{\sum_{s=1}^{m} x_i a_{si} p_s} + c_k^f \frac{\sum_{i=1}^{m}(1-\pi_i) x_i p_i + \sum_{i=1}^{m} p_i \sum_{j=1}^{m} a_{ij} \pi_j x_j}{\sum_{s=1}^{m} c_s^f p_s} + e_k \frac{\sum_{s=1}^{m} i_s p_s}{\sum_{s=1}^{m} e_s p_s} \leq x_k + i_k + \sum_{i=1}^{m} a_{ki} \pi_i x_i, \quad k = \overline{1,m}. \quad (29)$$

The next Theorem gives sufficient conditions for existence of economic equilibrium.

**Theorem 9.** Suppose that the non-negative vector $y = \{y_i\}_{i=1}^{m+2}$ solves the set of inequalities

$$\sum_{i=1}^{m} x_i a_{ki} y_i + c_k^f y_{m+1} + e_k y_{m+2} = x_k + i_k + \sum_{i=1}^{m} a_{ki} \pi_i x_i, \quad k \in I, \tag{30}$$

$$\sum_{i=1}^{m} x_i a_{ki} y_i + c_k^f y_{m+1} + e_k y_{m+2} < x_k + i_k + \sum_{i=1}^{m} a_{ki} \pi_i x_i, \quad k \in J, \tag{31}$$

and it is such that the Frobenius number of the matrix $A(y) = \left| \dfrac{a_{ij} y_j}{\pi_i} \right|_{i,j=1}^{m}$ equals one, where $I$ is non-empty subset of $M = \{1,2,...m\}$, $I \cup J = M$. If $p = \{p_i\}_{i=1}^{m}$ is such non-negative solution to the set of equations

$$\sum_{s=1}^{m} y_s a_{si} p_s = \pi_i p_i, \quad i = \overline{1,m}, \tag{32}$$

that

$$y_{m+1} = \frac{\sum_{i=1}^{m}(1-\pi_i) x_i p_i + \sum_{i=1}^{m} p_i \sum_{j=1}^{m} a_{ij} \pi_j x_j}{\sum_{s=1}^{m} c_s^f p_s}, \quad y_{m+2} = \frac{\sum_{s=1}^{m} i_s p_s}{\sum_{s=1}^{m} e_s p_s}, \quad \sum_{s=1}^{m} c_s^f p_s > 0, \sum_{s=1}^{m} e_s p_s > 0, \sum_{s=1}^{m} a_{si} p_s > 0, \ i = \overline{1,m},$$

then $p = \{p_i\}_{i=1}^{m}$ is an equilibrium price vector. If $J$ is empty subset, then in this case the demand equals supply.

The **Proof** of the Theorem is evident one. In fact, the vector $x + i + A\pi x$, where $\pi x = \{\pi_i x_i\}_{i=1}^{m}$, belongs to the interior of the cone generated by vector-columns of the matrix $C = |C_{ki}|_{k,i=1}^{m,m+2}$, where $C_{ki} = a_{ki} x_i$, $k, i = \overline{1,m}$, $C_{k,m+1} = c_k^f$, $k = \overline{1,m}$, $C_{k,m+2} = e_k$, $k = \overline{1,m}$, as the equality $Cy^0 = x + i + A\pi x$, holds where $y^0 = \{y_i^0\}_{i=1}^{m+2}$, $y_i^0 = 1 + \pi_i$, $i = \overline{1,m}$, $y_{m+1}^0 = y_{m+2}^0 = 1$. If the rank of the matrix $C$ equals $r \leq m$, then according to the Theorem 6.1.3 [2] there exists the set of such $m+2-r+1$ linearly independent non-negative solutions $y_r, y_{r+1},..., y_{m+2}$ to the set of equations $Cy = x + i + A\pi x$, with respect to the vector $y = \{y_i\}_{i=1}^{m+2}$ that arbitrary non-negative solution $y = \{y_i\}_{i=1}^{m+2}$ to this set of equations can be expressed in the form $y = \sum_{i=r}^{m+2} \gamma_i y_i$, where $\gamma_i \geq 0$, $i = \overline{r, m+2}$.

Therefore, the set of equations (30) has non-negative solution. And if it is such that inequalities (31) hold and the spectral radius of the matrix $A(y)$ equals one, then there exists non-zero price vector solving the set of equations (32) which is an equilibrium price vector. Let us note that components of equilibrium vector $p = \{p_i\}_{i=1}^{m}$, whose indices belong to the set $J$ vanish.

It is just the quality of equilibrium that determines how close is economy to recession. Multiplying by $p_k$ the $k$-th inequality and introducing notations $X_{ki} = p_k a_{ki} x_i$, $k, i = \overline{1,m}$, $X_k = x_k p_k$, $C_k^f = c_k^f p_k$,

$E_k = e_k p_k$, $I_k = i_k p_k$, $k = \overline{1,m}$, we can rewrite (29) in value terms as

$$\sum_{i=1}^{m} X_{ki} \frac{\pi_i X_i}{\sum_{s=1}^{m} X_{si}} + C_k^f \frac{\sum_{i=1}^{m}(1-\pi_i)X_i + \sum_{i=1}^{m}\sum_{j=1}^{m} X_{ij}\pi_j}{\sum_{s=1}^{m} C_s^f} + E_k \frac{\sum_{s=1}^{m} I_s}{\sum_{s=1}^{m} E_s} \leq X_k + I_k + \sum_{i=1}^{m} X_{ki}\pi_i, \quad k = \overline{1,m}. \quad (33)$$

**Definition 4.** We say that economy described aggregately is in an equilibrium state if inequalities (33) hold.

## 4. The Application to the Study of European Economies

In this Section, we apply the model explored in the previous Sections to the study of European economies. It is known that Ukrainian economy in 2010 was in recession. Below, we use the statistical data for such European countries as U K, Germany, Greece, Russia, and Ukraine to identify development trends of these countries. In contrast to [8], where the main parameters of the model are the macroeconomic indicators, namely, gross domestic product, household investment, consumption, and the recession is explained by the technological shock, we describe the recession phenomenon on the base on an equilibrium state quality accounting for the structure of production, consumption, investment, and price structure. In a recession, exchange mechanism breaking occurs manifested in the fact that in equilibrium the demand for goods, for which aggregate demand is less than supply, completely vanishes. Because of this, there is practically no purchase of this commodity group. The equilibrium prices for such a group of commodities can fall arbitrarily low. Devaluation of the national currency, the increase in unemployment, devaluation of deposits, and asset price fall is the result of all that. Let us denote demand vector by $D = \{D_k\}_{k=1}^m$, where

$$D_k = \sum_{i=1}^m X_{ki} \frac{\pi_i X_i}{\sum_{s=1}^m X_{si}} + C_k^f \frac{\sum_{s=1}^m (1-\pi_i) X_i + \sum_{i=1}^m \sum_{j=1}^m X_{ij}\pi_j}{\sum_{s=1}^m C_s^f} + E_k \frac{\sum_{s=1}^m I_s}{\sum_{s=1}^m E_s} - \sum_{i=1}^m X_{ki}\pi_i,$$

And the supply vector by $S = \{S_k\}_{k=1}^m$, where $S_k = X_k + I_k$. Then the vector of lacking demand can be written in the form $D - S = \{D_k - S_k\}_{k=1}^m$.

**Definition 5.** The $k$-th industry are said to create recession, if $D_k - S_k < 0$.

We present calculalation results for UK, Germany, Greece, Russia, and Ukraine in the form of vertically ordered histograms. We numbered industries upwards, putting right components of the vector $S = \{S_k\}_{k=1}^m$, and left such components of the vector $D - S = \{D_k - S_k\}_{k=1}^m$, for which $D_k - S_k < 0$.

To get an idea, how much a country is close to a recession we use quantitative characteristic for each country

$$r = \frac{\sum_{k:D_k-S_k<0} |D_k - S_k|}{\sum_{s=1}^m X_s - \sum_{i,j=1}^m X_{ij}}$$

being the ratio of total demand lacking to its gross domestic product. The calculations we have carried out show

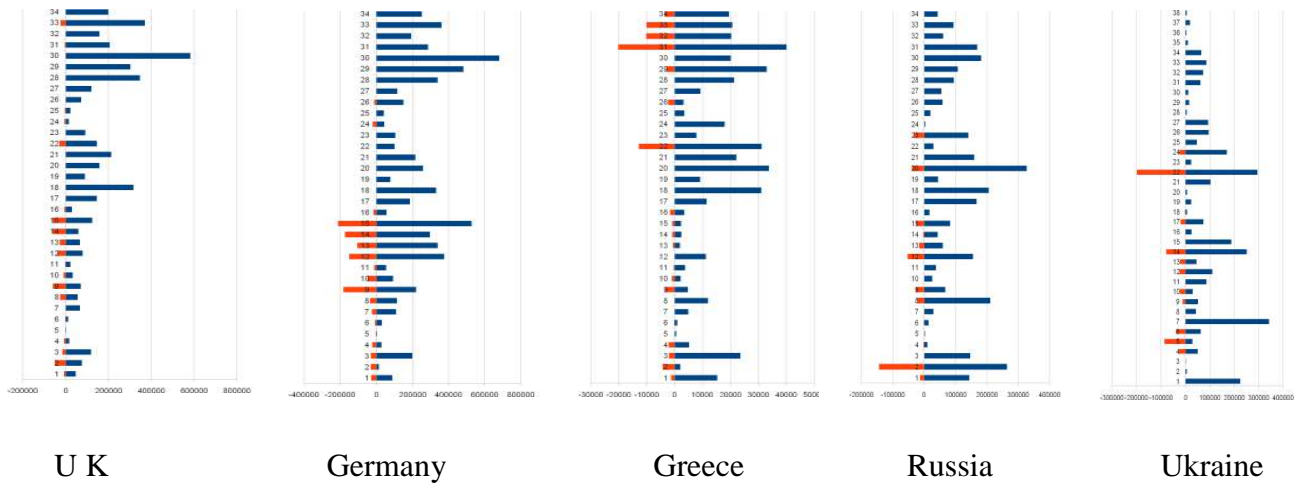

|   |   |   |   |   |
|---|---|---|---|---|
| U K | Germany | Greece | Russia | Ukraine |

Industries of U K, Germany, Greece, and Russia

| | | |
|---|---|---|
| 1 | Agriculture, Hunting, Forestry and Fishing |
| 2 | Mining and Quarrying |
| 3 | Food, Beverages and Tobacco |
| 4 | Textiles and Textile Products |
| 5 | Leather, Leather and Footwear |
| 6 | Wood and Products of Wood and Cork |
| 7 | Pulp, Paper, Paper , Printing and Publishing |
| 8 | Coke, Refined Petroleum and Nuclear Fuel |
| 9 | Chemicals and Chemical Products |
| 10 | Rubber and Plastics |
| 11 | Other Nonmetallic Mineral |
| 12 | Basic Metals and Fabricated Metal |
| 13 | Machinery, Nec |
| 14 | Electrical and Optical Equipment |
| 15 | Transport Equipment |
| 16 | Manufacturing, Nec; Recycling |
| 17 | Electricity, Gas and Water Supply |
| 18 | Construction |
| 19 | Sale, Maintenance and Repair of Motor Vehicles and Motorcycles; Retail Sale of Fuel |
| 20 | Wholesale Trade and Commission Trade, Except of Motor Vehicles and Motorcycles |
| 21 | Retail Trade, Except of Motor Vehicles and Motorcycles; Repair of Household Goods |
| 22 | Hotels and Restaurants |
| 23 | Inland Transport |
| 24 | Water Transport |
| 25 | Air Transport |
| 26 | Other Supporting and Auxiliary Transport Activities; Activities of Travel Agencies |
| 27 | Post and Telecommunications |
| 28 | Financial Intermediation |
| 29 | Real Estate Activities |
| 30 | Renting of M&Eq and Other Business Activities |

| 31 | Public Admin and Defence; Compulsory Social Security |
| 32 | Education |
| 33 | Health and Social Work |
| 34 | Other Community, Social and Personal Services |

r-coeff (Total Demand Reduction /Total Added Value)

| U K | 0.21 |
| Germany | 0.34 |
| Greece | 0.30 |
| Russia | 0.23 |
| Ukraine | 0.49 |

The 4 Most Sensitive Industries:

U K

| Demand Reduction | Gross Output | Import | Export | Number | Name |
| --- | --- | --- | --- | --- | --- |
| 52134 | 74188.45 | 57210.3 | 38142.76 | 2 | Mining and Quarrying |
| 1644.7 | 1674.369 | 1047.59 | 1508.7 | 5 | Leather, Leather and Footwear |
| 60014.8 | 70487.14 | 35400.1 | 64208.2 | 9 | Chemicals and Chemical Products |
| 59868 | 59072.3 | 41129.4 | 51231.69 | 14 | Electrical and Optical Equipment |

Germany

| Demand Reduction | Gross Output | Import | Export | Number | Name |
| --- | --- | --- | --- | --- | --- |
| 30607 | 13612.1 | 28372.38 | 6844.48 | 2 | Mining and Quarrying |
| 22445.23 | 28886. | 8585.4 | 28588.85 | 4 | Textiles and Textile Products |
| 3246.42 | 3833.98 | 1139.74 | 4274.92 | 5 | Leather, Leather and Footwear |
| 181654.1 | 219533.0 | 98310.1 | 185001 | 9 | Chemicals and Chemical Products |

Greece

| Demand Reduction | Gross Output | Import | Export | Number | Name |
| --- | --- | --- | --- | --- | --- |
| 4402.04 | 1942.37 | 7299.15 | 121.32 | 2 | Mining and Quarrying |
| 1032.09 | 4698.79 | 5758.15 | 1520.0 | 9 | Chemicals and Chemical Products |
| 1715.70 | 3360.40 | 345.23 | 142.8 | 16 | Manufacturing, Nec; Recycling |
| 2395.31 | 3201.1 | 4906.9 | 1878.57 | 26 | Other Supporting and Auxiliary Transport Activities; Activities of Travel Agencies |

Russia

| Demand Reduction | Gross Output | Import | Export | Number | Name |
| --- | --- | --- | --- | --- | --- |
| 143797.3 | 263447.40 | 858.82 | 167488.62 | 2 | Mining and Quarrying |
| 26971.37 | 67295.36 | 13469.93 | 22882.51 | 9 | Chemicals and Chemical Products |
| 53216. | 154878.18 | 9478.2 | 39146.53 | 12 | Basic Metals and Fabricated Metal |
| 26275.96 | 82260.89 | 35177.19 | 2550.97 | 15 | Transport Equipment |

Ukraine

| Demand Reduction | Gross Output | Import | Export | Number | Name |
| --- | --- | --- | --- | --- | --- |
| 86253.89 | 28986 | 109550 | 532 | 5 | Extraction of crude petroleum and natural gas |
| 38262.22 | 63386 | 11270 | 24554 | 6 | Mining of quarrying, except of energy producing materials |

| Demand Reduction | Gross Output | Import | Export | Number | Name |
|---|---|---|---|---|---|
| 21976.3 | 30038 | 809 | 3998 | 10 | Manufacture of coke oven products; processing of nuclear fuel |
| 199016.7 | 294920 | 880 | 428 | 22 | Trade; repair of motor vehicles, household appliances and personal demand items |

The 4 Most Significant Contributing Industries:

U K

| Demand Reduction | Gross Output | Import | Export | Number | Name |
|---|---|---|---|---|---|
| 52134 | 74188.45 | 57210.3 | 38142.76 | 2 | Mining and Quarrying |
| 60014.8 | 70487.14 | 35400.1 | 64208.2 | 9 | Chemicals and Chemical Products |
| 59868 | 59072.3 | 41129.4 | 51231.69 | 14 | Electrical and Optical Equipment |
| 63933.46 | 122624.0 | 35379.67 | 84809.0 | 15 | Transport Equipment |

Germany

| Demand Reduction | Gross Output | Import | Export | Number | Name |
|---|---|---|---|---|---|
| 181654.1 | 219533.0 | 98310.1 | 185001 | 9 | Chemicals and Chemical Products |
| 149993.28 | 376482.57 | 141124.87 | 167674.80 | 12 | Basic Metals and Fabricated Metal |
| 174238.3 | 296697.7 | 104912.26 | 205065.4 | 14 | Electrical and Optical Equipment |
| 212400.4 | 526518.06 | 75289.56 | 312487.9 | 15 | Transport Equipment |

Greece

| Demand Reduction | Gross Output | Import | Export | Number | Name |
|---|---|---|---|---|---|
| 12910.99 | 31153.14 | 8.34 | 11.09 | 22 | |
| 20263.99 | 40143.42 | 185.68 | 145.2 | 31 | Public Admin and Defence; Compulsory Social Security |
| 10283.52 | 20300.0 | 59.17 | 55.8 | 32 | Education |
| 10157.28 | 20687.45 | 69.66 | 86.95 | 33 | Health and Social Work |

Russia

| Demand Reduction | Gross Output | Import | Export | Number | Name |
|---|---|---|---|---|---|
| 143797.3 | 263447.40 | 858.82 | 167488.62 | 2 | Mining and Quarrying |
| 53216. | 154878.18 | 9478.2 | 39146.53 | 12 | Basic Metals and Fabricated Metal |
| 39594.2 | 325992.07 | 3854.12 | 76542.9 | 20 | Wholesale Trade and Commission Trade, Except of Motor Vehicles and Motorcycles |
| 33865. | 140558.29 | 4319.30 | 66208.79 | 23 | Inland Transport |

Ukraine

| Demand Reduction | Gross Output | Import | Export | Number | Name |
|---|---|---|---|---|---|
| 86253.89 | 28986 | 109550 | 532 | 5 | Extraction of crude petroleum and natural gas |
| 38262.22 | 63386 | 11270 | 24554 | 6 | Mining of quarrying, except of energy producing materials |
| 78621.4 | 251401 | 39995 | 143201 | 14 | |
| 199016.7 | 294920 | 880 | 428 | 22 | Trade; repair of motor vehicles, household appliances and personal demand items |

Ukrainian Industries

| # | Industry |
|---|---|
| 1 | Agriculture, hunting and related service activities |
| 2 | Forestry, logging and related service activities |
| 3 | Fishing, fish farming and related service activities |
| 4 | Mining of coal and lignite; extraction of peat; mining of uranium and thorium ores |
| 5 | Extraction of crude petroleum and natural gas |
| 6 | Mining of quarrying, except of energy producing materials |
| 7 | Manufacture of food products, beverages and tobacco |
| 8 | Manufacture of textiles and textile products; manufacture of wearing apparel; dressing and dyeing of fur |
| 9 | Manufacture of wood and wood products; manufacture of pulp, paper and paper products; publishing and printing |
| 10 | Manufacture of coke oven products; processing of nuclear fuel |
| 11 | Manufacture of refined petroleum products |
| 12 | Manufacture of chemicals and chemical products; manufacture of rubber and plastic products |
| 13 | Manufacture of other non-metallic mineral products |
| 14 | Manufacture of basic metals and fabricated metal products |
| 15 | Manufacture of machinery and equipment |
| 16 | Manufacturing n.e.c. |
| 17 | Production and distribution of electricity |
| 18 | Manufacture of gas; distribution of gaseous fuels through mains |
| 19 | Steam and hot water supply |
| 20 | Collection, purification and distribution of water |
| 21 | Construction |
| 22 | Trade; repair of motor vehicles, household appliances and personal demand items |
| 23 | Activity of hotels and restaurants |
| 24 | Activity of transport |
| 25 | Post and telecommunications |
| 26 | Financial activity |
| 27 | Real estate activities |
| 28 | Renting of machinery and equipment without operator and of personal and |
| 29 | Computer and related activities |
| 30 | Research and development |
| 31 | Other business activities |
| 32 | Public administration |
| 33 | Education |
| 34 | Health care and provision of social aid |
| 35 | Sewage and refuse disposal, sanitation and similar activities |
| 36 | Activities of membership organizations n.e.c. |
| 37 | Recreational, cultural and sporting activities |
| 38 | Other services activities |

Comparative analysis of the dynamic changes in a number of economies of European countries shows a clear tendency to recession manifestations in 2011. However, the features of their specific structures, especially focusing on strong export potential, often conceal recession pulses generated within national economies. For the quantitative comparison of recession trends in the economies of the European countries, it is reasonable to introduce the ratio of the demand decrease to wholesale national product. Although one should well understand that the scale invariance can not be here due to the structural differences of various economies, as well as lack of linear dependencies between intensive and extensive characteristics of economies, this option can serve as a quality characteristic of economic efficiency and stability of the system mainly as the result of economic management quality.

Quantitative analysis based on the formula (33) and the input-output balance [9] gave the following results. Ukraine's economy in 2010 was not in a state of equilibrium. In what follows, we give a drop in demand compared to the supply in millions of hryvnia (UAH), without specifying them near the numbers. Demand in value terms (millions UAH) fell compared to the supply in 14 industries of the Ukrainian economy.

| Industries | Decline | G D P | Import |
|---|---|---|---|
| Forestry + | 642 | 5 086 | |
| Coal + | 26 585 | 50 706 | 3 512 |
| Hydrocarbon | 86 254 | 28 986 | 109 950 |
| Minerals | 38 262 | 63 386 | 11 270 |
| Wood + | 11 304 | 51 454 | 16 379 |
| Coke + | 21 976 | 30 038 | 809 |
| Chemistry + | 23 072 | 224 366 | 110 223 |
| Metallurgy | 78 621 | 251 401 | 39 995 |
| Nonmetal mineral prods | 22 931 | 46 023 | 9 287 |
| Electricity | 19 917 | 73 520 | 48 |
| Gas + | 1 411 | 7 066 | 2 |
| Trade + | 199 020 | 294 920 | 880 |
| Transport | 24 607 | 170 415 | 2 |
| Real estate | 1 089 | 93 275 | 1 205 |

Thus, calculations on statistical data ground confirm that the Ukrainian economy in 2010 was in a deep recession due to falling demand for the main products. Abnormal drop in demand for gas production is due to the high import prices for this product. An important indicator of the fact that Ukraine's economy is in a recession is a significant demand decline in trade.

## 5. Conclusions

In the paper, we have proved the Theorems describing the structure of economic equilibrium states in the model of the exchange economy. For a given structure of demand vectors, we have studied the structure of property vectors under which given price vector is equilibrium one. We introduced the important notion of equivalent property distribution which was the ground for the proof that at the state of economic equilibrium there is such equivalent property distribution for which degeneracy multiplicity of equilibrium state is not less than the value declared in the Theorem 3.

In the Theorem 6, we described the structure of equilibrium states for which the demand for some group of goods is strictly less than the supply and which quantitatively characterized by decreasing real value of national currency. If the group of goods for which aggregate demand is strictly less than the supply becomes critical one, then national currency devalues, unemployment increases, bank deposits devalue, and asset values fall. Just this equilibrium state is the recession state.

In Theorems 7 and 8, we gave sufficient conditions for equilibrium existence at which the demand equals the supply. In the Theorem 9, we adapted the model of the state economy with production to the model considered containing sufficient conditions for equilibrium economy. In the last Section, we apply this model to analyze some European countries' economies.